\documentstyle[emulateapj,epsf]{article}

\def\etal{{\it et~al.}}
\def\oyobi{\&}

\begin{document}

\title{ General Relativistic Simulations of Jet Formation
in a Rapidly Rotating Black Hole Magnetosphere 
}

\author{Shinji Koide}
\affil{Faculty of Engineering, Toyama University, Gofuku 3190,
Toyama 930-8555, Japan}

\author{David L. Meier}
\affil{
Jet Propulsion Laboratory, 4800 Oak Grove Dr. Pasadena, CA 91109, USA.
}

\author{Kazunari Shibata} 
\affil{
Kwasan and Hida Observatories, Kyoto University, Yamashina,
Kyoto, 607-8471, Japan.
}

\author{Takahiro Kudoh}
\affil{
 National Astronomical Observatory, Mitaka, Tokyo 181-8588, Japan}

\begin{abstract}
    To investigate the formation mechanism of relativistic jets
    in active galactic nuclei and micro-quasars, we have
    developed a new general relativistic magnetohydrodynamic code in
    Kerr geometry.  Here we report on the first numerical simulation of
    jet formation in a rapidly-rotating
    ($a=0.95$) Kerr black hole magnetosphere.  
    We study cases in which the Keplerian accretion disk is both
    co-rotating and counter-rotating with respect to
    the black hole rotation.
    In the co-rotating disk case, our results are almost the same
    as those in Schwarzschild black hole cases: a gas pressure-driven 
    jet is formed by a shock in the disk,
    and a weaker magnetically-driven jet is also generated
    outside the gas pressure-driven jet.
    On the other hand, in the counter-rotating disk case,
    a new powerful magnetically-driven jet is formed inside
    the gas pressure-driven jet.
    The newly found magnetically-driven jet in the latter case is 
    accelerated by a strong 
    magnetic field created by frame dragging in the ergosphere.
    Through this process, the magnetic field extracts the energy 
    of the black hole rotation.
\end{abstract}

\vspace{0.5cm}

\noindent {\it Subject headings:} 
accretion, accretion disks --- black hole physics --- galaxies: jets
--- magnetic field --- methods: numerical --- MHD --- relativity

\section{Introduction}

    Radio jets ejected from radio loud active galactic nuclei (AGNs)
    sometimes show proper motion with apparent velocity exceeding the
    speed of light $c$ (\cite{pearson81,hughes91}).  
    The widely-accepted explanation for
    this phenomenon, called superluminal motion, is relativistic jet
    flow in a direction along the observer's line-of-sight with
    a Lorentz factor greater than 2 (\cite{rees66}).  
    Such relativistic motion is
    thought to originate from a region very close to the putative
    supermassive black hole which is thought to power each AGN 
    (\cite{lindenbell69,rees84}).  
    On the other hand, the great majority of AGNs are radio quiet and do not
    produce powerful relativistic radio jets (\cite{rees84}).  
    These two classes of
    active objects (radio loud and quiet) are also found
    in the black hole candidates (BHCs) in our own Galaxy.  Objects
    with superluminal jets, such as GRS 1915+105 and GRO J1655-40,
    belong to the radio loud class (\cite{mirabel94,tingay95}).  
    Other objects such as Cyg X-1 and GS 1124-68 are 
    relatively radio quiet and produce little or no jet.

    What causes the difference between the two classes?  Recent
    observations of the BHCs in our Galaxy suggest that the Galactic
    superluminal sources contain very rapidly rotating black holes 
    (normalized angular momentum, $a \equiv J/[GM_{\rm BH}^2/c]
    =0.9-0.95$, where $G$ and $M_{\rm BH}$ are the gravitational
    constant and black hole mass, respectively), 
    while the black holes in Cyg X-1 and GS 1124-68 are
    spinning much less rapidly ($a = 0.3-0.5$) (\cite{cui98}).  
    A similar rapidly rotating black hole is also suggested
    in the AGN of the Seyfert 1 galaxy MCG-6-30-15 by
    the X-ray satellite ASCA (\cite{iwasawa96}). According to
    recent (nonrelativistic) studies of magnetically-driven jets 
    from accretion disks by Kudoh \oyobi\ Shibata (1995, 1997a),
    the terminal velocity of the formed jet
    is comparable to the rotational velocity of the disk at the foot of
    the jet. Further nonrelativistic simulations of jet formation 
    confirm these results (\cite{kudo97b,ouyed97}), 
    except for the extremely large magnetic field/high jet-power case 
    (\cite{meier97,meier99})
    in which very fast jets can be produced.
    The rotation velocity at the innermost stable orbit of the
    Schwarzschild black hole ($r= 3r_S$) is $0.5c$, where
    $r_{\rm S} =2GM_{\rm BH}/c^2$ is the Schwarzschild radius.  
    In addition, it appears that the poloidal magnetic field
    strength in disks around non-rotating black holes may be 
    not extremely strong if the magnetic field energy density
    is comparable with that of the radiation (\cite{begelman84,rees84}).
    Therefore, a jet produced by MHD acceleration from an accretion
    disk around a non-rotating black hole should be sub-relativistic
    and very weak.  In fact, numerical simulations of jet formation in
    a Schwarzschild metric show only sub-relativistic jet flow (\cite{koide99}),
    except for the case when the initial black hole corona is in
    hydrostatic equilibrium rather than free fall
    (\cite{koide98}).

    Several mechanisms for relativistic jet formation from rotating
    black holes have been proposed 
    (\cite{blandford77,takahashi90}).  
    However, up until now no
    one has performed a self-consistent numerical simulation of the
    dynamic process of jet formation in a rotating black hole
    magnetosphere.  To this end, we have developed a Kerr
    general relativistic magnetohydrodynamic (KGRMHD) code.  In this
    paper we report briefly on what we believe are some
    of the first calculations of their kind --- 
    simulation of jet formation in a
    rotating black hole magnetosphere.

\section{Numerical Method}

We use a 3 + 1 formalism of the general relativistic conservation
laws of particle number, momentum, and energy and Maxwell
equations with infinite electric conductivity (\cite{thorne86}).
The Kerr metric, which describes the spacetime around a rotating 
black hole, is used in the calculation.
When we use Boyer-Lindquist coordinates,
$x^0=ct$, $x^1=r$, $x^2=\theta$, and $x^3=\phi$, the Kerr
metric $g_{\mu \nu}$ is written as follows,
\begin{equation}
ds^2=g_{\mu \nu}dx^{\mu}dx^{\nu}
=-h_0^2(cdt)^2+\sum_{i=1}^{3} h_i^2(dx^i)^2-2h_3\Omega _3 cdt dx^3 .
\end{equation}

\noindent By modifying the lapse function in our Schwarzschild black hole code 
($\alpha = \root \of {1-r_{\rm S}/r}$) to be 
$\alpha = \root \of {h_0^2+\Omega _3^2}$, 
and adding some terms of $\Omega _3$ to the time evolution equations,
we were able to develop a KGRMHD code relatively easily. 
(See Appendix C in \cite{koide99} for more details on this procedure 
and the meaning of symbols used.) 


We use the Zero Angular Momentum Observer (ZAMO) system
for the 3-vector quantities, such as velocity ${\bf v}$, magnetic field
${\bf B}$, and so on.
For scalars, we use the frame comoving with the fluid flow.
The simulation is performed
in the region $0.75r_{\rm S} \leq r \leq
20r_{\rm S}$, $0 \leq \theta \leq \pi /2$ with $210 \times 70$
mesh points, assuming axisymmetry with respect to the $z$-axis
and mirror symmetry with respect to the plane $z=0$.
A free boundary condition is employed
at $r=0.75 r_{\rm S}$ and $r=20r_{\rm S}$.
In the simulations, we use simplified tortoise coordinates,
$x={\rm log}(r/r_{\rm H} -1)$, where $r_{\rm H}$ is the 
radius of the black hole horizon. 
To avoid numerical oscillations, we use
a simplified TVD method (\cite{davis84,koide96,koide97,koide99}).
We checked the KGRMHD code by computing Kepler motion around
a rotating black hole and comparing with analytic results
(\cite{shapiro83}).

\section{Results}

The simulations were performed for two cases in which the disk co-rotates 
and counter-rotates with respect to the black hole rotation.
Figures 1a-c illustrate the time evolution of the 
counter-rotating disk case and
Fig. 1d the final state of the co-rotating case.
These figures show the rest mass density 
(color), velocity (vectors), and magnetic field (solid lines) in
$0 \leq R \equiv r {\rm sin} \theta
\leq 7 r_{\rm S}$, $0 \leq z \equiv r {\rm cos} \theta \leq 7 r_{\rm S}$.
The black region at the origin shows the inside of the black hole horizon.
The angular momentum parameter of the black hole is $a=0.95$
and the radius is $r_{\rm H} =0.656 r_{\rm S}$. 
The initial state in the simulation consists of
a hot corona and a cold accretion disk around the black hole (Fig. 1a).
In the corona, plasma is assumed to be in nearly
stationary infall, with the specific enthalpy
$h/\rho c^2 =1+\Gamma p/[(\Gamma -1)\rho c^2] =1.3$,
where $\rho$ is the rest mass density, $p$ is the pressure, and
$\Gamma$ is specific heat ratio and set $\Gamma =5/3$. 
Far from the hole,
it becomes the stationary transonic solution exactly.
The accretion disk is located at $|{\rm cot} \theta | \leq 0.125$,
$r \geq r_{\rm D} =3r_{\rm S}$ and the initial
velocity of the disk is assumed to be 
the velocity of a circular orbit around the Kerr black hole.
The co-rotating disk is stable,
but the counter-rotating disk is unstable
in the region $R \leq 4.4 r_{\rm S}$. Except for the disk rotation
direction, we use the same initial conditions in both cases.
The mass density of the disk is 100 times that of the
corona at the inner edge of the disk. 
The mass density profile is given by that of a
hydrostatic equilibrium corona with a scale height
of $r_{\rm c} \sim 3 r_{\rm S}$.
The disk is in pressure balance with the corona, and
the magnetic field lines are perpendicular to the accretion disk.
We use the azimuthal component of the vector potential $A_{\rm \phi}$
of the Wald solution to set the magnetic field, 
which provides a uniform magnetic field
far from the Kerr black hole (\cite{wald74}).
Here the magnetic field strength far from the black hole
is $0.3 \root \of {\rho _0 c^2}$, where $\rho _0$ is the
initial corona density at $r=3 r_{\rm S}$.
However, we do not use the time component of the 
vector potential $A_t$ from Wald solution; instead, 
we use the ideal MHD condition ${\bf E} +{\bf v} \times {\bf B} ={\bf 0}$
to determine the electric field ${\bf E}$.
Here the Alfv\'{e}n velocity and plasma beta value at the disk 
($r=3.5r_{\rm S}$) are $v_{\rm A} = 0.03c$ and $\beta \sim 3.4$,
respectively.

Figure 1b shows the state at $t= 30\tau _{\rm S}$,
where $\tau _{\rm S}$ is defined as $\tau _{\rm S} \equiv r_{\rm S}/c$.
By this time the inner edge of the disk has rotated $0.75$ cycles,
{\it if} we assume the edge is at $R = 3r_{\rm S}$.\footnote{To calculate 
inner disk rotation cycles, in this paper we 
always will assume that the inner edge is located at $R = 3r_{\rm S}$, 
regardless of how far inward the edge actually has accreted.}
Actually, the edge falls toward
the black hole and rotates faster at $R=2r_{\rm S}$.
The rapid infall produces a shock at $R = 3.1 r_{\rm S}$, and 
the high pressure behind it begins to produce the jet.
This is the same pressure-driven jet formation process seen previously 
in the Schwarzschild case (\cite{koide99}).

Figure 1c shows the final state of the counter-rotating disk case
at $t=47 \tau _{\rm S}$ when the inner edge of the disk
rotated 1.2 cycles.
The accretion disk continues to fall rapidly 
toward the black hole, with the disk plasma entering
the ergosphere and then crossing the horizon,
as shown by the crowded magnetic field lines near $r=0.75r_{\rm S}$.
The magnetic field lines become radial due to dragging by the disk
infall near the black hole.
The jet is ejected almost along the magnetic field lines.
Its maximum total and poloidal velocities are the same,
$v = v_{\rm p} =0.44c$ at 
$R=3.2r_{\rm S}$, $z=1.6r_{\rm S}$. 
The mass density plot 
(color) shows that the jet consists of
two layers. One is an inner, low density,
fast, magnetically-driven jet and the other is an outer, high density,
slow, gas pressure-driven jet. The latter comes from the disk near the shock
at $R =3.1 r_{\rm S}$ and is, therefore, similar 
to the gas pressure-driven jet of Koide, Shibata, \oyobi\ Kudoh (1998). 
The former is new and has never been seen in the Schwarzschild
black hole case.
It comes from the disk near the ergosphere, and is
accelerated as follows.
As there is no stable orbit at $R \leq 4.4 r_{\rm S}$,
the disk falls rapidly into the ergosphere.
Inside the static limit , the velocity of frame dragging
exceeds the speed of light ($c \Omega_3 /\alpha > c$), 
causing the disk to rotate 
in the {\em same} direction of the black hole rotation 
(relative to the fixed Boyer-Lindquist frame), even though it was 
initially counter-rotating. 
The rapid, differential frame dragging greatly enhances the 
azimuthal magnetic field, which then accelerates the flow upward 
and pinches it into a powerful collimated jet. 

Figure 1d shows a snapshot of the co-rotating disk case at
$t=47 \tau _{\rm S}$. The disk stops its 
infall near $R = 3 r_{\rm S}$
due to the centrifugal barrier with a shock at 
$r =3.4r_{\rm S}$. The high pressure behind the shock causes
a gas pressure-driven jet with total and poloidal velocities of
$v=v_{\rm p} =0.30c$
at $R=3.4r_{\rm S}$, $z=2.4r_{\rm S}$.
A detailed analysis shows that a weak magnetically-driven jet is
formed outside the gas pressure-driven jet with
maximum total and poloidal velocities of $v=0.42c$ and 
$v_{\rm p} =0.13c$, respectively. This two-layered shell
structure is similar to that of Schwarzschild black hole
case (\cite{koide98}).
The centrifugal barrier makes the disk take much long time
to reach the ergosphere, which causes the difference
between the co-rotating and counter-rotating disk cases.

To more fully illustrate the physics of the jet formation mechanism,
in figure 2 we show the plasma beta, $\beta \equiv p/(B^2/2)$ (color)
and the toroidal component of the magnetic field,
$B_\phi$ (contour) in the counter-rotating and
co-rotating disk cases at $t=47 \tau _{\rm S}$.
The blue color shows the region where 
magnetic field dominates the gas pressure;
light red---yellow shows where gas pressure is dominant;
and solid contour line shows negative azimuthal
magnetic field ($B_\phi <0$), while the broken line
the positive value ($B_\phi > 0$).
The toroidal component
of the magnetic field $B_{\phi}$ is negative and 
its absolute value is very large above the black hole in both cases.
The field increases to more than 10 times the initial magnetic field.
This amplification is caused by the shear of the plasma flow
in the Boyer-Lindquist frame
due to the frame dragging effect of the rotating black hole
(\cite{yokosawa91,yokosawa93,meier99}). 
Under the simplifying assumption that the plasma 
is at rest in the ZAMO frame,
the general relativistic Faraday law of induction and ideal MHD
condition yield,
\begin{equation}
\frac{\partial B_\phi}{\partial t} = f_1 B_r + f_2 B_\theta ,
\label{eqdyn}
\end{equation} 
where $f_1 = c(h_3/h_1) \partial (\Omega _3/h_3)/\partial r$,
and $f_2 = c(h_3/h_2) \partial (\Omega _3/h_3)/\partial \theta$.
This expression is almost identical to that of $\omega$-dynamo
effect from the field of the terrestrial magnetism.
Noted that $f_2$ is one order smaller than $f_1$ when $a \sim 1$.

Where does the magnetic field amplification energy 
comes from?
It does not come from the gravitational energy
or thermal energy of the disk, because frame dragging effect occurs even
when the plasmas of the disk and corona are rest and cool.
The only other possible energy source is the rotation of the
black hole itself. Indeed, the increase in the azimuthal magnetic field
component (eq. (\ref{eqdyn})) depends on the shear 
of the rotational variable $\Omega _3$.
We conclude that the amplification energy of the magnetic
field is supplied by extraction of the rotational energy
of the black hole. 



The distribution of the plasma beta ($\beta$)
and the azimuthal component of the magnetic field
($B_\phi$) of the counter-rotating and co-rotating disk cases
are quite different. In the co-rotating disk case,
they are similar to those of the Schwarzschild black hole
case. 
In the counter-rotating disk case, 
the outer part has a positive azimuthal component of the magnetic field
($B_\phi >0$), which is caused by the counter-rotating disk,
and the outer part has the high plasma beta. The inner part has a negative
azimuthal magnetic field ($B_\phi <0$) and low plasma beta.
Note that the very high plasma beta region (yellow region) 
is outside of the jet; at this point it has 
almost stopped and eventually will fall into the black hole.
The negative azimuthal magnetic field is caused by the
disk around the ergosphere, where the disk rotates in the same
direction as the black hole in the Boyer-Lindquist
frame.

To confirm the jet acceleration mechanism,
we estimate the power from the electromagnetic field,
$W_{\rm EM}={\bf v} \cdot ({\bf E} + {\bf J} \times {\bf B}$)
and the gas pressure, $W_{\rm gp} = - {\bf v} \cdot \nabla p$
along the line, $z=1.1r_{\rm S}$ which crosses the jet foot
(Fig. 3).
At $t=47 \tau _{\rm S}$, the gas pressure is dominant
in the co-rotating disk case (Fig. 3b).
However, in the counter-rotating disk case, the 
electromagnetic power is dominant near the black hole even through
the gas pressure power is the same as that 
of the co-rotating disk case (Fig. 3a).
The magnetically-driven jet in this latter case
is accelerated by the magnetic field anchored to the 
ergospheric disk. The frame dragging effect rapidly
rotates the disk in the same direction 
as the black hole rotation, increasing the
azimuthal component of the magnetic field and
the magnetic tension which, in turn, accelerates the plasma by the 
magnetic pressure and centrifugal force, respectively. 
(A detailed analysis shows that both component of the
magnetic forces are comparable.)
This mechanism of jet production, therefore, is a kind of Penrose 
process that uses the magnetic field to extract rotational energy of 
the black hole and eject a collimated outflow from very near the 
horizon.

\section{Discussion}

We have presented general relativistic simulations
of jet formation from both counter-rotating and co-rotating
disks in a Kerr black hole magnetosphere.
We have found that jets are formed in both cases.
At the time when the simulations 
were stopped ($t=47 \tau _{\rm S}$ ($53 \tau _{\rm S}$),
after the inner edge of the disk had rotated 1.2 (1.4) cycles
in the counter-rotating (co-rotating) disk case)
the poloidal velocities of the jets were $v \sim 0.4c$ (counter-rotating),
$\sim 0.3c$ (co-rotating), both sub-relativistic.
In the co-rotating disk case, the jet has a two-layered
structure: inner, gas pressure-driven jet and outer,
magnetically-driven jet. On the other hand, in the counter-rotating
case, a new magnetically-driven jet has been found
inside the gas pressure-driven jet. The new jet is accelerated
by the magnetic field induced by the frame dragging
effect in the ergosphere. In this case, existence of a 
magnetically-driven jet is not clear outside the gas pressure-driven
jet. A longer term simulation may show a three-layered structure
including the outer, magnetically-driven jet.

Unfortunately, the 
counter-rotating (co-rotating) disk case
could not be continued beyond $t=47 \tau _{\rm S}$ 
($t=53 \tau _{\rm S}$) because of 
numerical problems.  
We have performed one other case previously --- the infall of a 
magnetized non-rotating disk into a rapidly-rotating black hole
(\cite{koide99b}).  
The disk falls toward the black hole more rapidly than
the counter-rotating case.
At later times (after almost two inner 
disk turns) it developed a {\em relativistic}
jet with a velocity of $v \sim 0.9c$ (Lorentz factor $\sim 2$). 
We believe that, if we had been able to perform longer-term simulations 
here, in at least the counter-rotating disk case the 
magnetically-driven jet 
also would have been accelerated to relativistic velocities (and possibly the 
co-rotating case as well).
Despite its low speed, the magnetically-driven jet in 
the counter-rotating disk case is nevertheless noteworthy 
because it extracts rotational energy from the black hole. 
While the process is similar to the Blandford-Znajek mechanism
(\cite{blandford77}), it appears more closely related to the model
of Takahashi \etal\ (1990). 
In our case, the electromagnetic field energy is transformed
immediately into kinetic energy in the jet.  
A more detailed analysis and further calculations will be
reported in our next paper.

Recently, Wardle \etal\ (1998) detected circularly
polarized radio emission from the jets of the 
archetypal quasar 3C279. They concluded
that electron-positron pairs are
important components of the jet plasma.
Similar detections in three other radio sources have been
made (\cite{homan99}), which suggests that, in general,
extragalactic radio jets may be composed mainly
of an electron-positron pair plasma.
The electron-positron plasma is probably produced very near
the black hole. The mechanism we have investigated here,
magnetically-driven jet powered by 
extraction of rotational energy of a black hole,
is a strong candidate for explaining the acceleration of
such electron-positron jets.

\vspace{1.0cm} 


S. K. thanks M. Inda-Koide for discussions
and important comments for this study. 
We thank K.-I. Nishikawa, M. Takahashi, A. Tomimatsu, P. Hardee,
and J.-I. Sakai
for discussions and encouragement. 
We appreciate the support of the National Institute for Fusion Science
and the National Astronomical Observatory
in the use of their super-computers.
Part of this research was carried out at the
Jet Propulsion Laboratory, California Institute of Technology,
under contract with the National Aeronautics and
Space Administrations.

\bibliographystyle{alpha}

\begin{thebibliography}{99}

\bibitem[Begelman, Blandford, \oyobi\ Rees 1984]{begelman84} 
Begelman, .M.C., Blandford, R.D., Rees, M.J. 1984,
Review of Modern Physics, {\bf 56}, 255.

\bibitem[Belloni \etal\ 1997]{belloni97b} 
Belloni, T., van der Klis, M., Lewin, W. H. G., van Paradijs, J.,
Dotani, T., Mitsuda, K., \oyobi\ Miyamoto, S. 1997, A \oyobi\ A {\bf 322},
857.

\bibitem[Blandford \oyobi\ Payne 1982]{blandford82} 
Blandford, R. D. \oyobi\ Payne, D. G. 1982, MNRAS
{\bf 199}, 883.

\bibitem[Blandford \oyobi\ Znajek 1977]{blandford77} 
Blandford, R. D. \oyobi\ Znajek, R. 1977, MNRAS
{\bf 179}, 433.

\bibitem[Cui, Zhang, \oyobi\ Chen 1998]{cui98}
Cui, W., Zhang, S. N., \oyobi\ Chen, W. 1998, 
ApJ {\bf 484}, 383.

\bibitem[Cui \etal\ 1997]{cui97}
Cui, W., Zhang, S. N., Focke, W., \oyobi\ Swank, J. H. 1997, 
MNRAS {\bf 290}, L65.

\bibitem[Davis 1984]{davis84}
Davis, S. F. 1984, NASA Contractor Rep. 172373,
ICASE Rep. No. 84-20.

\bibitem[Homan \etal\ 1999]{homan99}
Homan, D. C., Wardle, J. F. C., Roberts, D. H., and
Ojha, R. 1999, preprint.

\bibitem[Hughes 1991]{hughes91} 
Hughes, P. A. 1991, eds., {\it Beams and Jets in 
Astrophysics} (Cambridge University Press, New York).

\bibitem[Iwasawa \etal\ 1996]{iwasawa96} 
Iwasawa, K., Fabian, A. C., Reynolds, C. S.,
Nandra, K., Otani, C. Inoue, H. Hayashida, K. Brandt, W. N.,
Dotani, T., Kunieda, H., Matsuoka, M., Tanaka, Y. 1996, 
MNRAS, {\bf 282}, 1038.

\bibitem[Koide 1997]{koide97} 
Koide, S. 1997, ApJ {\bf 478}, 66. 

\bibitem[Koide, Nishikawa \oyobi\ Mutel 1996]{koide96} 
Koide, S., Nishikawa, K.-I., \oyobi\ Mutel, R. L. 1996, 
ApJ {\bf 463}, L71. 

\bibitem[Koide, Shibata, \oyobi\ Kudoh 1998]{koide98} 
Koide, S., Shibata, K., \oyobi\ Kudoh, T. 1998, ApJ {\bf 495}, L63. 

\bibitem[Koide, Shibata, \oyobi\ Kudoh 1999a]{koide99} 
Koide, S., Shibata, K., \oyobi\ Kudoh, T., 1999a, ApJ
{\bf 522}, 175.

\bibitem[Koide, Meier, Shibata, \oyobi\ Kudoh 1999b]{koide99b}
Koide, S., Meier, D. L., Shibata, K., \oyobi\  Kudoh, T. 1999b,
in Proc. 19th Texas Symp. on Relativistic Astrophysics,
ed. E. Aubourg, T. Montmerle, \oyobi\ J. Paul 
(World Scientific Press, Paris) in press.

\bibitem[Kudo \oyobi\ Shibata 1995]{kudo95} 
Kudoh, T. \oyobi\ Shibata, K. 1995, 
ApJ {\bf 452}, L41. 

\bibitem[Kudo \oyobi\ Shibata 1997a]{kudo97a}
Kudoh, T. \oyobi\ Shibata, K. 1997a, ApJ {\bf 474}, 362.

\bibitem[Kudo \oyobi\ Shibata 1997b]{kudo97b} 
Kudoh, T. \oyobi\ Shibata, K. 1997b, 
ApJ {\bf 476}, 632.

\bibitem[Linden-Bell 1969]{lindenbell69}
Linden-Bell, D. 1969, Nature {\bf 223}, 690.


\bibitem[Meier 1999]{meier99}
Meier, D. L. 1999, ApJ {\bf 522}, in press.

\bibitem[Meier \etal\ 1997]{meier97} 
Meier, D. L., Edgingon, S., Godon, P., Payne, D. G., \oyobi\ Lind, K. R. 1997,
Nature {\bf 388}, 350.

\bibitem[Mirabel \oyobi\ Rodriguez 1994]{mirabel94} 
Mirabel, I. F. \oyobi\ Rodriguez, L. F. 1994, 
Nature {\bf 371}, 46. 

\bibitem[Morgan \etal\ 1997]{morgan97}
Morgan, E. H., Remillard, R. A., \oyobi\ Greiner, J. 1977, ApJ
{\bf 482}, 993.

\bibitem[Ouyed, Pudritz, \oyobi\ Stone 1997]{ouyed97} 
Ouyed, R., Pudritz, R. E., \oyobi\ Stone, J. M. 1997, 
Nature {\bf 385}, 409. 
      
\bibitem[Pearson \etal\ 1981]{pearson81} 
Pearson, J. J. 1981, \etal\ Nature {\bf 290}, 365.


\bibitem[Rees 1966]{rees66} 
Rees, M. J. 1966, Nature {\bf 211}, 468. 

\bibitem[Rees 1984]{rees84} 
Rees, M. J. 1984, Ann. Rev. Astron. Ap. {\bf 22}, 471. 

\bibitem[Remillard \etal\ 1997]{remillard97}
Remillard, R. A, Morgan, E. H, McClintock, J. E,
Bailyn, C. D., Oroszek, A., \oyobi\ Greiner, J. 1999, 
in Proc. 18th Texas Symp. on Relativistic Astrophysics,
ed. A. Olinto, J. Frieman, \oyobi\ D. Schramm 
(World Scientific Press, Singapore) in press.

\bibitem[Shapiro \oyobi\ Teukolsky 1983]{shapiro83} 
Shapiro, S. L. \oyobi\ Teukolsky, S. A. 1983, 
{\it Black Holes, White Dwarfs, and Neutron Stars},
(John Wiley \oyobi\ Sons Inc., New York)

\bibitem[Shibata \oyobi\ Uchisa 1986]{shibata86} 
Shibata, K. \oyobi\ Uchida, Y. 1986, 
PASJ {\bf 38}, 631. 

\bibitem[Takahashi \etal\ 1990]{takahashi90} 
Takahashi, M., Nitta, S., Tatematsu, Y., \oyobi\ Tomimatsu, A. 1990, 
ApJ {\bf 363}, 206.

\bibitem[Thorne, Price, \oyobi\ Macdonald 1986]{thorne86}
Thorne, K. S., Price, R. H., \oyobi\ Macdonald, D. A. 1986, 
{\it Membrane Paradigm} (Yale University Press, New Haven and 
London).

\bibitem[Tingay \etal\ 1995]{tingay95} 
Tingay, S. J. 1995, \etal\ Nature {\bf 374}, 141.


\bibitem[Uchisa \oyobi\ Shibata 1985]{uchida85} 
Uchida, Y. \oyobi\ Shibata, K. 1985, 
PASJ {\bf 37}, 515. 

\bibitem[Wald 1974]{wald74}
Wald, R. M. 1974, Phys. Rev. D {\bf 10}, 1680. 

\bibitem[Wardle \etal\ 1998]{wardle98}
Wardle, J. F. C., Homan, D. C., Ojha, R., \oyobi\
Roberts, D. H. 1998, Nature {\bf 395}, 457.

\bibitem[Yokosawa 1993]{yokosawa93}
Yokosawa, M. 1993, PASJ {\bf 45}, 207.

\bibitem[Yokosawa, Ishizuka, \oyobi\ Yabuki 1991]{yokosawa91}
Yokosawa, M., Ishizuka, T., \oyobi\ Yabuki, Y. 1991, 
PASJ {\bf 43}, 427.



\end{thebibliography}


\newpage
\noindent {\Large Figure Captions}

\vspace{1cm}

\begin{list}{}{\setlength{\leftmargin}{0.5cm} \setlength{\itemindent}
{-0.5cm}}
\item
Fig. 1.
Time evolution of jet formation in the counter-rotating
disk case and the final state of the co-rotating disk case.
Color shows the logarithm of the proper mass density;
vectors indicate velocity; solid lines show the
poloidal magnetic field.
The black fan-shaped region at the origin shows the horizon of the
Kerr black hole ($a=0.95$). The dashed line near the horizon is
the inner boundary of the calculation region.

At $t=0$ and $t=30 \tau _{\rm S}$ 
the state of the co-rotating and counter-rotating disk 
cases are almost identical.  
However, at $t=47 \tau _{\rm S}$, while the infall of the disk 
in the co-rotating disk stops
(due to a centrifugal barrier), 
the unstable orbits of the counter-rotating disk plasma
continue to spiral rapidly toward the black hole horizon.
This difference causes the magnetohydrodynamic jet 
formation mechanisms in the two cases to differ drastically, resulting 
in a powerful jet emanating from deep within the ergosphere.

\item
Fig. 2.
Plasma beta (color) and azimuthal component of the
magnetic field $B_\phi$ (contour) in the counter-rotating
and co-rotating disk cases. Solid lines show 
negative value of $B_\phi$ and the dashed lines positive 
values.  The two cases differ 
significantly in structure, with the jet in the counter-rotating 
case originating much closer to the black hole.

\item
Fig. 3.
Power contribution to jet acceleration 
along the line, $z=1.1 r_{\rm S}$
due to the gas pressure ($W_{\rm gp}$)
and the electromagnetic force ($W_{\rm EM}$) 
for both the counter-rotating and co-rotating disk cases.
The jet in the counter-rotating disk
case is accelerated mainly by electromagnetic forces,
while that in the co-rotating disk is accelerated 
mainly by gas pressure.
Note that, while the power in the gas jet component is 
comparable in the two cases, the power in the MHD jet component is 
nearly two orders of magnitude greater in the counter-rotating case 
than the co-rotating case.  
\end{list}

\begin{figure}
\epsscale{1}
\plotone{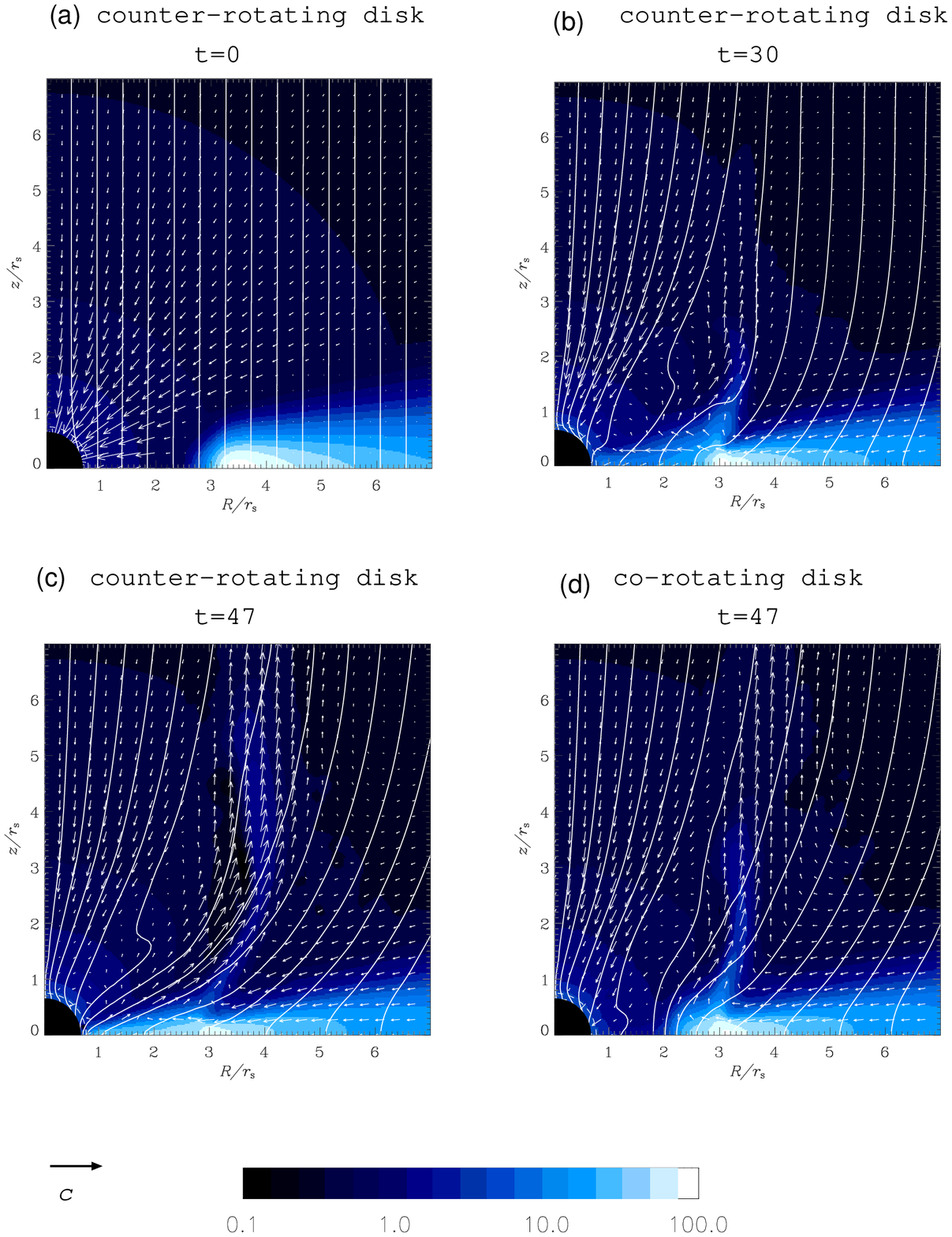}

\vspace{1cm}

Fig. 1.---
\end{figure}

\begin{figure}
\epsscale{1}
\plotone{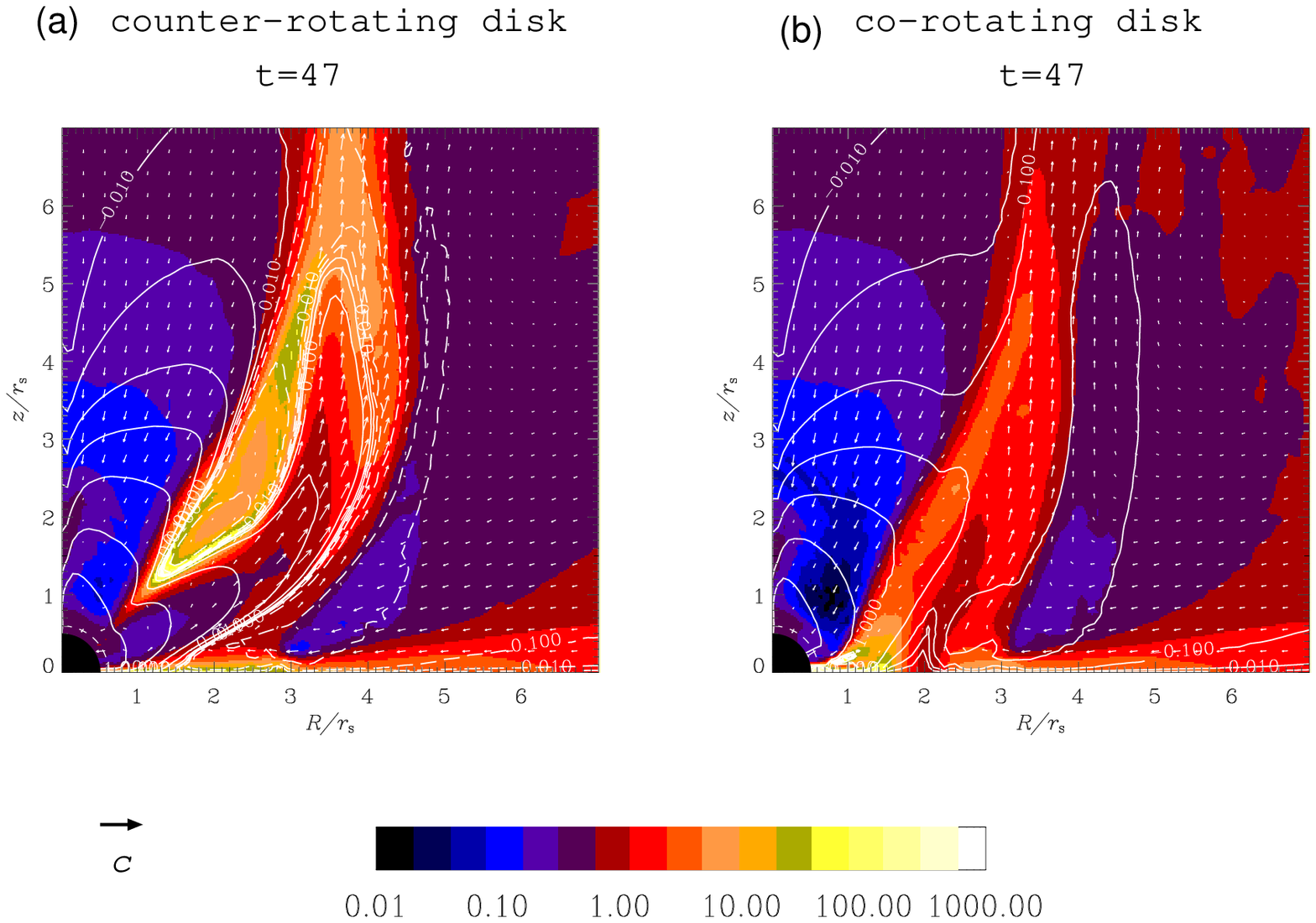}

\vspace{1cm}

Fig. 2.---
\end{figure}

\begin{figure}
\epsscale{1}
\plotone{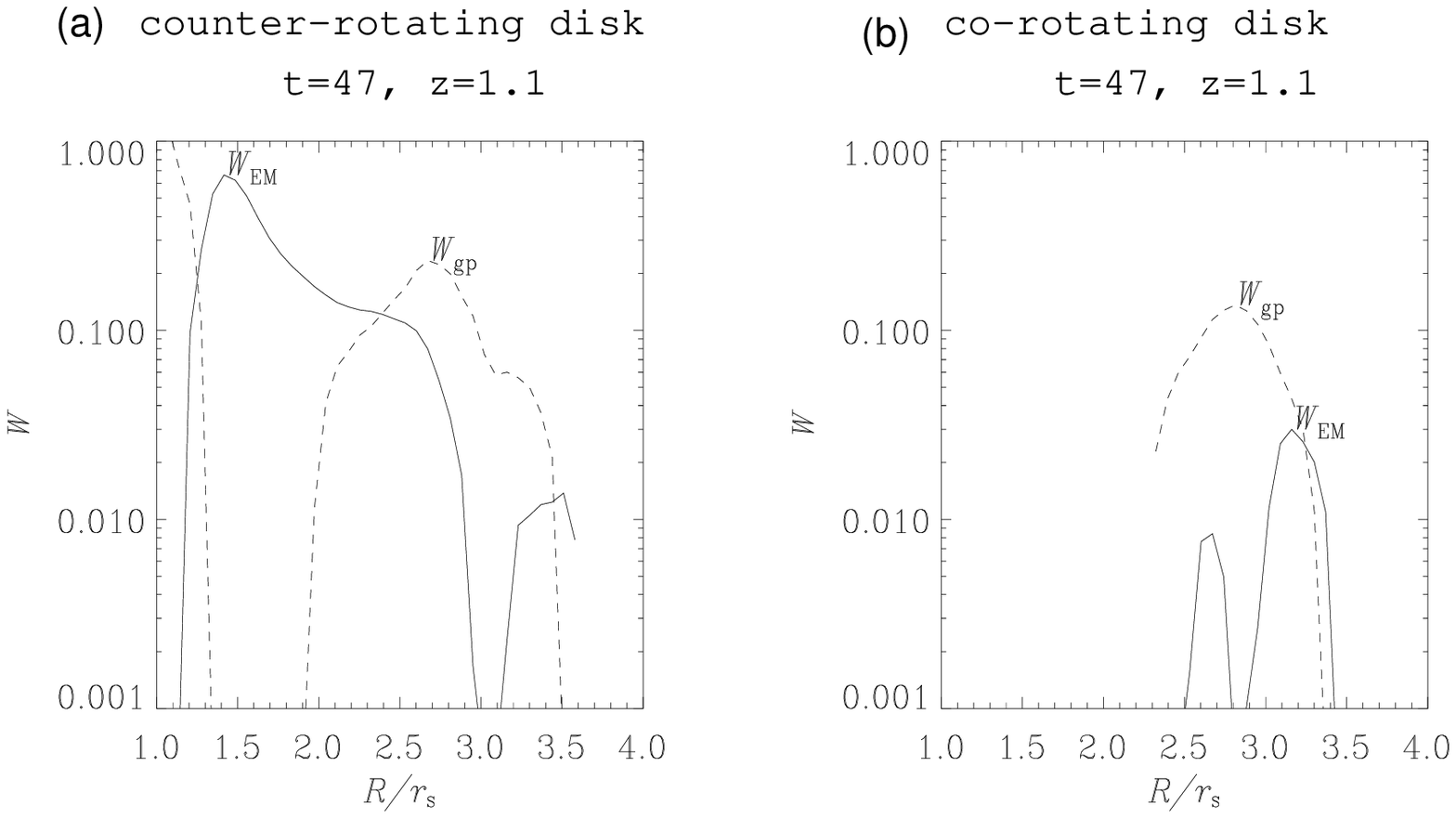}

\vspace{1cm}

Fig. 3.---
\end{figure}
\end{document}